\documentclass[aps,pra,reprint,superscriptaddress,amsmath,amssymb,floatfix]{revtex4-2}

\usepackage{graphicx}  
\usepackage{dcolumn}  
\usepackage{bm}     
\usepackage{hyperref}  
\usepackage{physics}
\usepackage[version=4]{mhchem} 
\usepackage{multirow}
\usepackage{longtable}
\usepackage{array}
\usepackage[normalem]{ulem}
\usepackage{makecell}

\hypersetup{
    colorlinks=true,
    linkcolor=blue,
    filecolor=magenta,      
    urlcolor=blue,
    citecolor=blue,
}

\begin{document}

\title{Laser cooling of \ce{Th^-}: Revised bound-state structure\\
and a more efficient cooling scheme}

\author{Shaowei Tian}
\affiliation{Shanghai EBIT Lab, Key Laboratory of Nuclear Physics and Ion-beam Application, Institute of Modern Physics, Fudan University, Shanghai 200433, China}

\author{Chaofan Shi}
\affiliation{Shanghai EBIT Lab, Key Laboratory of Nuclear Physics and Ion-beam Application, Institute of Modern Physics, Fudan University, Shanghai 200433, China}

\author{Ran Si}
\email{rsi@fudan.edu.cn}
\affiliation{Shanghai EBIT Lab, Key Laboratory of Nuclear Physics and Ion-beam Application, Institute of Modern Physics, Fudan University, Shanghai 200433, China}

\author{Chuangang Ning}
\affiliation{Department of Physics, State Key Laboratory of Low-Dimensional Quantum Physics, Tsinghua University, Beijing 100084, China}

\author{Chongyang Chen}
\email{chychen@fudan.edu.cn}
\affiliation{Shanghai EBIT Lab, Key Laboratory of Nuclear Physics and Ion-beam Application, Institute of Modern Physics, Fudan University, Shanghai 200433, China}

\begin{abstract}
The thorium anion ($\text{Th}^-$) has been proposed as a candidate for laser cooling of negative ions. However, previous theoretical and experimental results exhibit significant discrepancies in both the excitation energies and lifetimes. Here we revisit the bound-state structure and transition properties of $\text{Th}^-$ using two independent \textit{ab initio} approaches. The two calculations agree well and improve agreement with measured excitation energies and lifetimes. Based on the present level structure, we reassign several observed levels and show that the previously proposed $^4F^e_{3/2}\leftrightarrow{}^2S^o_{1/2}$ cooling transition is not closed and requires two repumping lasers. We therefore propose an alternative cooling scheme based on the $^4F^e_{3/2}\leftrightarrow{}^4F^o_{3/2}$ transition, which requires only one repumping laser and has a transition rate approximately five times larger ($A\approx1.1\times10^5~\mathrm{s}^{-1}$), the fastest reported bound--bound E1 transition in atomic anions to date.
\end{abstract}

\maketitle

\textit{Introduction.} Laser cooling has enabled major advances in quantum many-body physics~\cite{Bloch2008,Greiner2002}, precision spectroscopy and metrology~\cite{Ludlow2015,Takamoto2005,Rosenband2008}, and quantum information processing~\cite{Cirac1995,Monroe1995}, and has been successfully implemented for both neutral atoms and positive ions~\cite{Wineland1978,Neuhauser1978,Chu1985,Raab1987}. However, extending laser cooling to negative ions (anions) poses distinct challenges. Unlike neutral atoms or positive ions, the outermost electron in an anion is bound primarily by short-range electron correlation and polarization effects~\cite{Andersen2004, Pegg2004}. As a result, atomic anions typically possess only a few bound states and often lack bound excited states of opposite parity~\cite{Ning2022,Andersen1999}, which are necessary for strong electric-dipole (E1) transitions and efficient laser cooling. Once a suitable anion is laser cooled, other negatively charged particles can in principle be cooled sympathetically~\cite{Larson1986}, particularly antiprotons~\cite{Kellerbauer2006,Cerchiari2018La,Tang2021}, which could facilitate precision tests of CPT symmetry and antimatter gravity~\cite{Borchert2022,Ahmadi2018,Anderson2023}.

To date, bound--bound E1 transitions have been established in only four atomic anions: $\text{Ce}^-$~\cite{Walter2007,Walter2011,Fu2020Ce}, $\text{Os}^-$~\cite{Bilodeau2000,Kellerbauer2006,Warring2009,Fischer2010,OMalley2010,Pan2010,Lu2023Os},  $\text{La}^-$~\cite{OMalley2010,Pan2010,Walter2014,Jordan2015,Kellerbauer2015,Cerchiari2018La,Lu2019La,Walter2025La}, and $\text{Th}^-$~\cite{Tang2019,Tang2021,Zhang2025}. These anions face different challenges in realizing an efficient cooling cycle. \ce{Ce-} has numerous low-lying metastable states that may interrupt the optical cycle, while \ce{Os-} has a relatively slow E1 transition rate, with $\tau_{\rm exp}=201(10)~\mu\mathrm{s}$ for its cooling transition~\cite{Lu2023Os}. \ce{La-} has a much stronger E1 transition, $\tau_{\rm exp}=20.4(2.1)~\mu\mathrm{s}$~\cite{Cerchiari2018La}, but its only stable isotope, $^{139}\mathrm{La}$, has a nuclear spin of $I=7/2$, leading to hyperfine splitting and multiple repumping frequencies~\cite{Jordan2015,Cerchiari2018La}. By contrast, $^{232}\mathrm{Th}$ has $I=0$ and no hyperfine splitting, while \ce{Th^-} also has a comparatively strong candidate cooling transition with $\tau_{\rm exp}=30(2)~\mu\mathrm{s}$~\cite{Zhang2025}. These properties make \ce{Th^-} a particularly promising candidate for laser cooling.

The laser cooling potential of \ce{Th^-} was first established through measurements and calculations of its electron affinity and bound-state structure, leading to the proposal of the ${}^4F^e_{3/2}\leftrightarrow{}^2S^o_{1/2}$ transition as a candidate cooling transition~\cite{Tang2019}. This bound--bound E1 transition was subsequently observed at $2428.4(6)~\mathrm{nm}$ using resonant two-photon detachment spectroscopy~\cite{Tang2021}. More recently, three additional bound states were observed, and the lifetime of the proposed ${}^2S^o_{1/2}$ upper cooling state was measured to be $30(2)~\mu\mathrm{s}$~\cite{Zhang2025}, substantially shorter than the previous theoretical value of $85.5~\mu\mathrm{s}$~\cite{Tang2019}. Significant discrepancies were also found in the excitation energies. Most notably, the lowest odd-parity state was observed at $1847(13)~\mathrm{cm}^{-1}$~\cite{Zhang2025}, whereas the previous calculation predicted only $401~\mathrm{cm}^{-1}$~\cite{Tang2019}. 

Accurate theoretical calculations for \ce{Th^-} are particularly challenging because of its complex open-shell and strongly correlated electronic structure, making a comprehensive treatment of electron correlation essential. However, previous calculations~\cite{Tang2019,Tang2021} included only a limited treatment of electron correlation, which was insufficient to accurately describe the many-electron wave functions. Recent advances in computational methods and software implementations make substantially larger-scale atomic structure calculations feasible~\cite{Si2025Graspg,Cheung2025pCI}, enabling a more comprehensive reassessment of the bound-state structure and transition properties of \ce{Th^-}.

In this Letter, we investigate the bound-state structure of \ce{Th^-} and its laser cooling schemes using the configuration interaction plus all-order (CI+all-order) and multiconfiguration Dirac-Hartree-Fock plus relativistic configuration interaction (MCDHF+RCI) methods. The two independent calculations predict the same number of bound states and almost the same level ordering, while showing better agreement with available experimental energies and lifetimes. Based on the present level structure and reported photodetachment channels, we propose revised assignments for several observed levels. We show that the previously proposed ${}^4F^e_{3/2}\leftrightarrow{}^2S^o_{1/2}$ cooling transition does not form a closed optical cycle, contrary to earlier claims~\cite{Tang2019,Tang2021,Zhang2025}, and requires two additional repumping lasers. We therefore propose an alternative scheme based on the ${}^4F^e_{3/2}\leftrightarrow{}^4F^o_{3/2}$ transition, which requires only one repumping laser and has a transition rate about five times larger, enabling faster cooling while reducing experimental complexity.

\textit{Theoretical methods.} \ce{Th^-} has 91 electrons, and the electronic configurations of its bound states are $[\mathrm{Rn}]6d^3 7s^2$ for even parity and $[\mathrm{Rn}]6d^2 7s^2 7p$ for odd parity~\cite{Tang2019,Tang2021,Zhang2025} (the $[\mathrm{Rn}]$ core is omitted below for brevity). The CI+all-order method combines the configuration interaction (CI) method with the linearized coupled-cluster single-double (all-order) method~\cite{Safronova2009}. In this method, the five electrons outside the $[\mathrm{Rn}]$ core are treated as valence electrons by CI, which accounts for valence-valence (VV) correlations, while core-valence (CV) and core-core (CC) correlations are included by the all-order method. The many-electron wave function is expanded as a linear combination of Slater determinants, $\Psi=\sum_i c_iD_i$. The energies and wave functions of the low-lying states are obtained by diagonalizing the effective Hamiltonian,
\begin{equation}
H_{\mathrm{eff}}(E_n)\Psi_n=E_n\Psi_n,
\quad
H_{\mathrm{eff}}(E_n)=H+\Sigma(E_n),
\end{equation}
where $H$ is the Hamiltonian in the frozen-core approximation and $\Sigma(E_n)$ is an energy-dependent correction operator obtained from the all-order treatment of core excitations~\cite{Safronova2009}.

Treating \ce{Th^-} as a five-valent system with a closed $[1s^2,\ldots,6p^6]$ core, we first solved the Dirac-Hartree-Fock equations for the core electrons and then constructed the $7s$, $6d$, $5f$, $7p$, $8s$, $8p$, $5g$, $7d$, and $6f$ orbitals in the frozen-core potential. The one-electron basis set was constructed using 40 B-splines of order 7 within a spherical cavity of radius $R=80$ a.u., including orbitals up to $n=35$ for seven partial waves ($l_{\max}=6$). The Breit interaction was included throughout the construction. This basis set was sufficiently large for the all-order calculation to construct an accurate effective Hamiltonian for the subsequent CI diagonalization.

After constructing the basis set, the next step is to construct an appropriate CI space that is computationally affordable yet sufficiently large to capture the strong electron correlation effect in \ce{Th^-}. We therefore examined two complementary strategies for CI space construction: active set expansion and iterative multireference (MR) expansion. The active set expansion enlarges the orbital space for single and double (SD) valence excitations, mainly improving the description of dynamical correlations. In contrast, the iterative MR expansion incorporates important near-degenerate configurations and provides an efficient description of static correlations.

Convergence tests demonstrate that further enlargement of the active set yields only a modest improvement in the excitation energies at a disproportionately large computational cost, whereas the MR expansion achieves better agreement with experiment at lower computational cost, indicating the importance of static correlations associated with strong configuration mixing in the low-lying states of \ce{Th^-}. Therefore, a sufficiently extended MR space but a limited active set were adopted for the final CI calculation, which contains approximately $4.9\times10^{7}$ even-parity and $5.0\times10^{7}$ odd-parity Slater determinants. Detailed CI space construction strategies and convergence tests are provided in the Supplemental Material~\cite{SM}.

In addition to the CI+all-order method, an independent MCDHF+RCI approach~\cite{Fischer2016} was employed using the GRASPG package~\cite{Si2025Graspg}, an optimized version of GRASP2018~\cite{Fischer2019Grasp2018}. In this method, the atomic state function (ASF) is expanded as a linear combination of configuration state functions (CSFs), $\Psi_J=\sum_i c_i\Phi_i$. In the MCDHF step, the radial orbitals and mixing coefficients were optimized self-consistently based on the Dirac-Coulomb Hamiltonian. The CSF expansions were generated from an MR set by allowing SD excitations of the five valence electrons to an active space enlarged layer by layer, with the orbitals from the preceding layers kept fixed and only the newly added orbitals optimized. Based on the convergence tests, the $10sp9d7fg$ active space was adopted for the subsequent RCI calculations.

After the MCDHF procedure, large-scale RCI calculations were performed with the radial orbitals fixed and with the Breit interaction and leading QED corrections included in the Hamiltonian. In this step, we adopted a strategy similar to the MR expansion used in the CI+all-order calculations.
The MR set was enlarged based on the mixing coefficients obtained from the MCDHF calculations. The RCI expansion was then generated from the enlarged MR set by retaining SD excitations of the five valence electrons and including SD excitations from the $6s$ and $6p$ subshells and single excitations from the $5d$ and $5p$ subshells, thereby extending the treatment of VV correlations and incorporating additional CV and CC correlations.
% \rsi{The MR set was further enlarged, guided by the mixing coefficients obtained from the MCDHF calculations, not only to improve the description of the VV correlations but also of the CV and CC correlations—specifically, the CV and CC correlations involving the $6s$ and $6p$ subshells, as well as the CV correlations involving the $5d$ and $5p$ subshells were included.}
% \sout{The MR set was further enlarged according to the configuration weights obtained from the MCDHF calculations  
% to improve the description of VV correlations and strong configuration mixing. In addition, SD excitations from the $6s$ and $6p$ subshells and single excitations from the $5d$ and $5p$ subshells were included to explicitly account for the relevant core correlations. }

The inclusion of these extensive electron correlation effects results in a prohibitively large RCI space of approximately $2.3\times10^9$ CSFs, distributed over different $J^\pi$ symmetries. We therefore employed the zero-first (ZF) method~\cite{Li2025Blockwise} together with CSFG-based condensation~\cite{Si2025Graspg,Li2023Reducing} to reduce the computational cost while retaining the significant correlation contributions. After this procedure, approximately $4.9\times10^8$ CSFs were retained for the final RCI calculations. This large-scale calculation provides a more comprehensive description of electron correlation effects in \ce{Th^-} than previous MCDHF+RCI calculations~\cite{Tang2019,Tang2021}. Detailed convergence tests for the active space, MR set, core correlations, and ZF treatment are provided in the Supplemental Material~\cite{SM}.

\textit{Bound-state structure.} Table~\ref{tab:energy_levels} summarizes the calculated excitation energies of the bound states of \ce{Th^-}, together with previous theoretical results and available experimental values. The spectroscopic term labels used in the present calculations are based on the leading $LS$ components of the MCDHF+RCI wave functions and should be regarded only as approximate identifiers because of strong configuration mixing.

\begin{table}[htbp]
\caption{\label{tab:energy_levels}
Theoretical and experimental excitation energies of \ce{Th^-}, in cm$^{-1}$. The two ``Level'' columns list the assignments adopted in previous and present works, respectively. Methods I and II denote the CI+all-order and MCDHF+RCI calculations, respectively. Previous results are taken from Refs.~\cite{Tang2019,Tang2021,Zhang2025}.
}
\renewcommand{\arraystretch}{1.0}

\begin{ruledtabular}
\begin{tabular}{@{}ccccccc@{}}

\multicolumn{3}{c}{Previous work} &
\multicolumn{3}{c}{This work} & Previous work \\
\cline{1-3}\cline{4-6} \cline{7-7}

\multirow{2}{*}{Level}
& \multirow{2}{*}{Theo.}
& \multirow{2}{*}{Expt.}
& \multirow{2}{*}{Level}
& \multicolumn{2}{c}{Theo.}
& \multirow{2}{*}{Expt.} \\
\cline{5-6}

& & & & I & II & \\
\hline
\noalign{\vspace{2pt}}
\multicolumn{7}{l}{$6d^3 7s^2$} \\
\noalign{\vspace{2pt}}
$^4F^e_{3/2}$ & 0    & 0          & $^4F^e_{3/2}$ & 0    & 0    & 0          \\
$^4F^e_{5/2}$ & 1377 & 1657(6)    & $^4F^e_{5/2}$ & 1603 & 1641 & 1657(6)    \\
$^4F^e_{7/2}$ & 2642 & 2892.9(50) & $^4F^e_{7/2}$ & 2844 & 2916 & 2892.9(50) \\
$^4F^e_{9/2}$ & 3637 & 3166.8(59) & $^4P^e_{3/2}$ & 3257 & 3286 & 3166.8(59)\footnotemark[1] \\
$^4P^e_{3/2}$ & 4284 & --         & $^4P^e_{1/2}$ & 3663 & 3762 &  --  \\
$^4P^e_{1/2}$ & 4940 & --         & $^4F^e_{9/2}$ & 3665 & 3814 & 3666(12)\footnotemark[1]\footnotemark[2] \\

\hline
\noalign{\vspace{2pt}}
\multicolumn{7}{l}{$6d^2 7s^2 7p$} \\
\noalign{\vspace{2pt}}
$^4G^o_{5/2}$ & 401  & 1847(13)   & $^4G^o_{5/2}$ & 1681 & 1715 & 1847(13)   \\
$^4F^o_{3/2}$ & 3033 & --         & $^4F^o_{3/2}$ & 3570 & 3666 & --         \\
$^2S^o_{1/2}$ & 3904 & 4118.0(10) & $^2S^o_{1/2}$ & 4019 & 4015 & 4118.0(10) \\
$^4F^o_{7/2}$ & 3974 & 3666(12)   & $^4G^o_{7/2}$ & 4433 & 4470 & --         \\
$^4F^o_{5/2}$ & 3992 & 4592.6(10) & $^2D^o_{3/2}$ & 4502 & 4449 & 4592.6(10)\footnotemark[1] \\
$^4D^o_{3/2}$ & 4445 & --         & $^2D^o_{5/2}$ & 4504 & 4465 & 4618.1(10)\footnotemark[1] \\
$^4D^o_{1/2}$ & 4503 & 4618.1(10) &                 &      &      &            \\

\end{tabular}
\end{ruledtabular}
\footnotetext[1]{Previous assignments of these experimental levels are revised.}
\footnotetext[2]{This level is tentatively assigned to $^4F^e_{9/2}$; $^4P^e_{1/2}$ cannot be excluded.}
\end{table}

The independent CI+all-order and MCDHF+RCI calculations yield nearly identical bound-state structures below the measured electron affinity threshold, $\mathrm{EA}=4901.35(48)~\mathrm{cm}^{-1}$~\cite{Tang2019}. The excitation energies obtained by the two methods have a mean absolute relative difference of only $1.8\%$, with a standard deviation of $1.1\%$. Both methods predict a total of twelve bound states, consisting of six even-parity states and six odd-parity states. In contrast, the bound-state structure predicted by previous theoretical studies~\cite{Tang2019,Tang2021} differs substantially from the present results. We attribute these differences mainly to the more complete treatment of electron correlation in this work, including more extensive core correlations and significantly enlarged MR sets that provide a better description of the strongly correlated electronic structure of \ce{Th^-}.

Based on the present bound-state structure, we reanalyzed the experimental measurements. These measurements provide the excitation energies of the observed levels but do not uniquely determine their total angular momenta or spectroscopic term assignments. Consequently, the assignments adopted in previous studies relied largely on comparisons between the measured energies and the earlier theoretical level structure. Since the present calculations predict a different level structure, some of the previous assignments should be reconsidered. The following discussion is given in order of increasing excitation energy.

The previous assignments of the observed levels at $1657$, $1847$, and $2892.9~\mathrm{cm}^{-1}$ as $^4F^e_{5/2}$, $^4G^o_{5/2}$, and $^4F^e_{7/2}$, respectively, are retained. The present calculations show significantly improved agreement with experiment, particularly for the lowest odd-parity $^4G^o_{5/2}$ state. The previous theoretical excitation energy of this state was only $401~\mathrm{cm}^{-1}$, lower than the experimental value by a factor of approximately $4.6$, whereas the present results agree with experiment to within $9.0\%/7.1\%$. In the following discussion, paired values are given as CI+all-order/MCDHF+RCI unless otherwise stated.

The assignment of the level at $3166.8~\mathrm{cm}^{-1}$ requires reconsideration. Previous work assigned this level to the previously closest calculated $^4F^e_{9/2}$ state, with an excitation energy of $3637~\mathrm{cm}^{-1}$, approximately $15\%$ higher than experiment. In the present calculations, the $^4P^e_{3/2}$ state becomes the closest level, with excitation energies of $3257/3286~\mathrm{cm}^{-1}$, showing much better agreement with the observed value. Moreover, photodetachment from this level to the neutral Th~${}^3F^e_2$, ${}^3F^e_3$, and ${}^3P^e_2$ states has been observed, with the ${}^3F^e_2$ and ${}^3P^e_2$ channels exhibiting stronger relative intensities than the ${}^3F^e_3$ channel~\cite{Zhang2025}. This observation further supports the reassignment, since the $^4P^e_{3/2}$ state possesses more allowed $p$-wave photodetachment channels to $^3F^e_2$ and $^3P^e_2$ than to the ${}^3F^e_3$ state, whereas the opposite tendency is expected for the ${}^4F^e_{9/2}$ state. We therefore propose that the level at $3166.8~\mathrm{cm}^{-1}$ be reassigned to $^4P^e_{3/2}$.

The observed level at $3666~\mathrm{cm}^{-1}$, previously assigned to the $^4F^o_{7/2}$ state, is difficult to assign unambiguously. In the present calculations, $^4F^o_{7/2}$ is no longer predicted to be bound, while the calculated $^4P^e_{1/2}$, $^4F^e_{9/2}$, and $^4F^o_{3/2}$ states are all close to the observed level. However, $^4F^o_{3/2}$ has a very short lifetime, making it unlikely to be observed in the photodetachment spectrum. Both $^4P^e_{1/2}$ and $^4F^e_{9/2}$ are relatively long-lived and allow $p$-wave photodetachment to the observed neutral Th ${}^3F^e_{2,3}$ states~\cite{Zhang2025}. The observed stronger ${}^3F^e_3$ channel is consistent with more allowed channels from $^4F^e_{9/2}$ to ${}^3F^e_3$ than to ${}^3F^e_2$, whereas $^4P^e_{1/2}$ shows the opposite tendency. We therefore tentatively reassign this level to $^4F^e_{9/2}$, although $^4P^e_{1/2}$ cannot be excluded.

Three resonances T1--T3 were observed at $4118.0$, $4592.6$, and $4618.1~\mathrm{cm}^{-1}$, respectively, in Ref.~\cite{Tang2021}. The previous assignment of T1 to the ${}^2S^o_{1/2}$ state is retained, with improved agreement with experiment, whereas the assignments of T2 and T3 require reconsideration. They were previously tentatively assigned to the $^4F^o_{5/2}$ and $^4D^o_{1/2}$ states, respectively. However, under these assignments, the reported $\ce{Th^-}\,{}^4F^o_{5/2}\rightarrow \mathrm{Th}\,{}^3P^e_0$ and $\ce{Th^-}\,{}^4D^o_{1/2}\rightarrow \mathrm{Th}\,{}^3F^e_3$ channels~\cite{Tang2021} would proceed through $d$-wave photodetachment, while the $s$-wave channels are forbidden by angular momentum conservation. Since higher partial waves are strongly suppressed near threshold according to the Wigner threshold law, these assignments are less favorable. In the present calculations, the ${}^2D^o_{3/2}$, ${}^2D^o_{5/2}$, and ${}^4G^o_{7/2}$ states are found near these two resonances, but ${}^4G^o_{7/2}$ is excluded because the E1 transition from the ${}^4F^e_{3/2}$ ground state is forbidden. We therefore reassign T2 and T3 to ${}^2D^o_{3/2}$ and ${}^2D^o_{5/2}$, respectively, which is the only assignment allowing all reported channels to proceed through $s$-wave photodetachment.

% As a further consistency check, we examined the photodetachment channels for all experimentally observed levels. For each assignment proposed in the present work, all the reported channels can proceed through $s$- or $p$-wave photodetachment, without involving $d$ waves or higher-order partial waves. This behavior is consistent with the expected suppression of higher-order partial waves near threshold and therefore provides additional support for these assignments.

In addition to accurate excitation energies, reliable transition properties are essential for evaluating the laser cooling schemes. We therefore calculated the line strengths ($S$), transition rates ($A$), branching fractions (BFs), and radiative lifetimes ($\tau$) using both the CI+all-order and MCDHF+RCI methods. Only transitions with BFs above $10^{-7}$ are considered, and the detailed properties of these transitions are provided in the Supplemental Material~\cite{SM}. Fig.~\ref{fig:s_comparison_e1_m1} compares the $S$ values obtained by the two methods. Since $S$ is independent of the transition energy, this comparison directly reflects the agreement between the calculated transition matrix elements.

\begin{figure}[htbp]
    \centering
\includegraphics[width=\columnwidth]{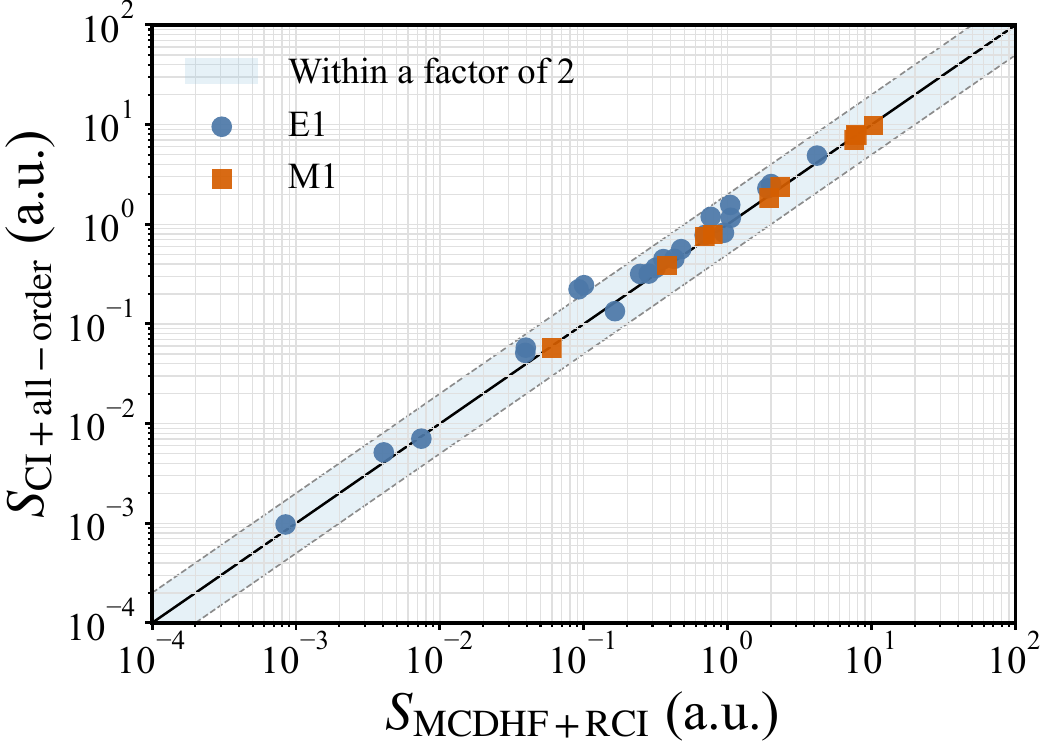}
\caption{Comparison of the line strengths calculated by the CI+all-order and MCDHF+RCI methods. The dashed line indicates perfect agreement, and the shaded region denotes agreement within a factor of 2.}
    \label{fig:s_comparison_e1_m1}
\end{figure}

Overall, the two methods show good consistency in the calculated transition properties, with 30 of the 32 line strengths in Fig.~\ref{fig:s_comparison_e1_m1} agreeing within a factor of 2. Similar agreement is found for the transition rates, branching fractions, and radiative lifetimes, with detailed results provided in the Supplemental Material~\cite{SM}. The calculated lifetimes can be further compared with experiment. For the ${}^2S^o_{1/2}$ state, the calculated lifetimes are $41.3/48.1~\mu\mathrm{s}$, in reasonable agreement with the experimental value of $30(2)~\mu\mathrm{s}$~\cite{Zhang2025}, and much closer than the previous theoretical value of $85.5~\mu\mathrm{s}$~\cite{Tang2019}. For the reassigned ${}^2D^o_{3/2}$ and ${}^2D^o_{5/2}$ states, the calculated lifetimes are $25.1/38.2~\mu\mathrm{s}$ and $85.6/97.0~\mu\mathrm{s}$, respectively, compared with the measured values of $62(5)$ and $53(3)~\mu\mathrm{s}$~\cite{Zhang2025}.

\textit{Laser cooling schemes.} The revised bound-state structure has direct implications for the laser cooling schemes of \ce{Th^-}. The ${}^4F^e_{3/2} \leftrightarrow {}^2S^o_{1/2}$ transition has previously been proposed as a potential laser cooling transition~\cite{Tang2019}. The present calculations give transition rates of $A=2.40\times10^4/2.07\times10^4~\mathrm{s}^{-1}$, corresponding to the Doppler cooling limits of $T_D=0.092/0.079~\mu\mathrm{K}$. One-dimensional cooling of an ensemble of \ce{Th^-} ions from $100~\mathrm{K}$ to the Doppler limit would require approximately $8.7\times10^4$ photon scattering cycles, corresponding to ideal cooling times of $7.2/8.4~\mathrm{s}$ under two-level saturation conditions. The main challenge in realizing laser cooling on this transition is maintaining a sufficiently closed optical cycle over approximately $10^5$ photon scattering events.

\begin{figure}[t]
    \centering
    \includegraphics[width=0.9\columnwidth]{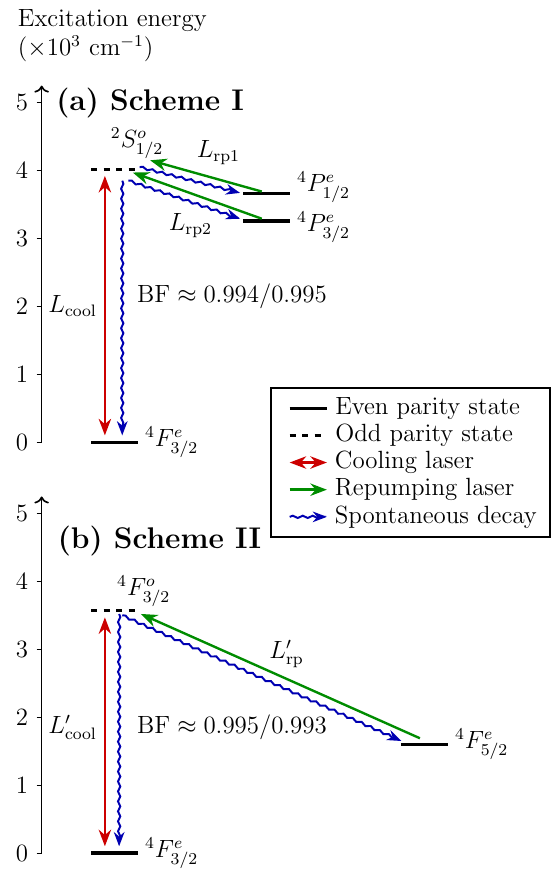}
\caption{Proposed laser cooling and repumping schemes for \ce{Th^-}.
(a) Scheme I is based on the $^4F^e_{3/2}\leftrightarrow{}^2S^o_{1/2}$ cooling transition and requires two repumping lasers for the ${}^4P^e_{1/2}$ and ${}^4P^e_{3/2}$ states. 
(b) Scheme II is based on the $^4F^e_{3/2}\leftrightarrow{}^4F^o_{3/2}$ cooling transition and requires only one repumping laser for the $^4F^e_{5/2}$ state. 
}
    \label{fig:cooling_scheme}
\end{figure}

However, as shown in Fig.~\ref{fig:cooling_scheme}(a), the revised bound-state structure places both the ${}^4P^e_{3/2}$ and ${}^4P^e_{1/2}$ states below the ${}^2S^o_{1/2}$ upper cooling state, opening additional E1 decay channels with BFs of $5.89\times10^{-3}/4.66\times10^{-3}$ and $6.08\times10^{-4}/2.24\times10^{-4}$, respectively. Although weak, these channels continuously remove population from the main cooling cycle. Since the ${}^4P^e_{3/2}$ ($\tau=468/535~\mathrm{ms}$) and ${}^4P^e_{1/2}$ ($\tau=12.4/12.3~\mathrm{s}$) states are metastable, the accumulated population cannot return efficiently to the cooling cycle through spontaneous decay. To maintain the cooling cycle, the population accumulated in the ${}^4P^e_{3/2,1/2}$ states must be repumped back to the ${}^2S^o_{1/2}$ state, as shown in Fig.~\ref{fig:cooling_scheme}(a). Based on the present bound-state structure, the two repumping transitions lie in the mid- to far-infrared region. Consequently, in contrast to the previously considered closed cooling cycle, this scheme requires two additional repumping lasers, substantially increasing the experimental complexity.

We therefore consider an alternative laser cooling scheme based on the $^4F^e_{3/2}\leftrightarrow{}^4F^o_{3/2}$ transition, as illustrated in Fig.~\ref{fig:cooling_scheme}(b). The calculated transition energies are $3570/3666~\mathrm{cm}^{-1}$, corresponding to wavelengths of $2.80/2.73~\mu\mathrm{m}$. The ${}^4F^o_{3/2}\rightarrow{}^4F^e_{3/2}$ cooling transition has BFs of $0.9954/0.9934$, while the main leakage channel to the metastable ${}^4F^e_{5/2}$ state ($\tau=6.96/6.39~\mathrm{s}$) has BFs of $4.58\times10^{-3}/6.60\times10^{-3}$. A much weaker leakage channel to the ${}^4P^e_{3/2}$ state is also present, with BFs of only $3.37\times10^{-5}/2.66\times10^{-5}$. Population leaking to ${}^4P^e_{3/2}$ predominantly returns to the ground state through the ${}^4G^o_{5/2}$ state. Multilevel rate equation calculations further show that additional repumping of ${}^4P^e_{3/2}$ reduces the estimated cooling time by only $14.9\%/9.8\%$ (see Supplemental Material~\cite{SM}). Given this limited improvement and the additional experimental complexity, we propose repumping only ${}^4F^e_{5/2}$ through the ${}^4F^e_{5/2}\leftrightarrow{}^4F^o_{3/2}$ transition, with calculated wavelengths of $5.08/4.94~\mu\mathrm{m}$. 
% Because this leakage is extremely weak and the leaked population decays predominantly back to the ${}^4F^e_{3/2}$ ground state through the ${}^4G^o_{5/2}$ state, no additional repumping laser is required for this channel. Thus, only the ${}^4F^e_{5/2}$ state needs to be repumped, and the cooling cycle can be closed by a single repumping laser on the ${}^4F^e_{5/2}\leftrightarrow{}^4F^o_{3/2}$ transition, with calculated wavelengths of $5.08/4.94~\mu\mathrm{m}$.

Thus, compared with the ${}^4F^e_{3/2}\leftrightarrow{}^2S^o_{1/2}$ scheme, this alternative scheme requires only one cooling laser and one repumping laser, making it experimentally more straightforward. Moreover, the calculated transition rates of the ${}^4F^e_{3/2}\leftrightarrow{}^4F^o_{3/2}$ cooling transition are $A=1.13\times10^5/1.04\times10^5~\mathrm{s}^{-1}$. These rates are approximately five times those of the ${}^4F^e_{3/2}\leftrightarrow{}^2S^o_{1/2}$ cooling transition and enable substantially faster cooling. This makes it the fastest reported bound--bound E1 transition in atomic anions to date. The corresponding Doppler cooling limits are estimated to be $0.43/0.40~\mu\mathrm{K}$. One-dimensional cooling of \ce{Th^-} ions from $100~\mathrm{K}$ to the Doppler limit would require approximately $9.7\times10^4/9.5\times10^4$ photon scattering cycles, corresponding to ideal cooling times of $1.7/1.8~\mathrm{s}$ under two-level saturation conditions.

\textit{Conclusions.} We have revisited the bound-state structure of \ce{Th^-} and its laser cooling schemes using two independent relativistic many-body approaches, the CI+all-order and MCDHF+RCI methods. The two calculations give consistent bound-state structures and improved agreement with experiment. Based on the present bound-state structure and reported photodetachment channels, we propose revised assignments for several observed levels.
The revised bound-state structure has direct implications for laser cooling of \ce{Th^-}. In particular, the previously proposed ${}^4F^e_{3/2}\leftrightarrow{}^2S^o_{1/2}$ transition does not form a closed cooling cycle and requires two additional repumping lasers. We therefore propose an alternative scheme based on the ${}^4F^e_{3/2}\leftrightarrow{}^4F^o_{3/2}$ transition, which requires only one repumping laser and has a transition rate about five times larger, enabling substantially faster cooling while retaining a sub-$\mu$K Doppler limit.
Overall, these results provide a more reliable description of the bound-state structure of \ce{Th^-} and a more experimentally accessible scheme for its laser cooling, establishing \ce{Th^-} as one of the most promising candidates for laser cooling of negative ions.

\textit{Acknowledgments.} This work is supported by the National Key Research and Development Project of China (No. 2022YFA1602303 and No. 2022YFA1602500) and the National Natural Science Foundation of China (Nos. 12674316, 12393284, 12374244, and 12341401). The authors thank Charles Cheung for helpful discussions.

\nocite{Stathopoulos1994DVDSON,Kozlov2015CIMBPT}
\bibliography{ref} 

\end{document}

\title{Supplemental Material for\\
Laser cooling of \ce{Th^-}: Revised bound-state structure\\
and a more efficient cooling scheme}

\author{Shaowei Tian}
\affiliation{Shanghai EBIT Lab, Key Laboratory of Nuclear Physics and Ion-beam Application, Institute of Modern Physics, Fudan University, Shanghai 200433, China}

\author{Chaofan Shi}
\affiliation{Shanghai EBIT Lab, Key Laboratory of Nuclear Physics and Ion-beam Application, Institute of Modern Physics, Fudan University, Shanghai 200433, China}

\author{Ran Si}
\email{rsi@fudan.edu.cn}
\affiliation{Shanghai EBIT Lab, Key Laboratory of Nuclear Physics and Ion-beam Application, Institute of Modern Physics, Fudan University, Shanghai 200433, China}

\author{Chuangang Ning}
\affiliation{Department of Physics, State Key Laboratory of Low-Dimensional Quantum Physics, Tsinghua University, Beijing 100084, China}

\author{Chongyang Chen}
\email{chychen@fudan.edu.cn}
\affiliation{Shanghai EBIT Lab, Key Laboratory of Nuclear Physics and Ion-beam Application, Institute of Modern Physics, Fudan University, Shanghai 200433, China}

\maketitle

\section{Computational Details}
\label{sec:computational_details}

\subsection{CI+all-order calculations}
\label{sec:sm_ci_all_order}

The CI+all-order calculations were performed with the pCI package~\cite{Cheung2025pCI}. We replaced the original Davidson eigensolver in pCI with an optimized restarted block-Davidson solver based on the DVDSON library used in GRASP~\cite{Stathopoulos1994DVDSON}. In our benchmarks, this modification accelerated the diagonalization by more than an order of magnitude and improved the stability of multi-root diagonalizations.

We examined two strategies for constructing the configuration interaction (CI) space: active set expansion and iterative multireference (MR) expansion. The even-parity calculation is used below to illustrate these strategies; the same procedures were also applied to the odd-parity calculation. In the active set expansion, we allowed all possible single and double (SD) valence excitations from $6d^3 7s^2$, $6d^3 7s 8s$, and $6d^2 7s^2 7d$ to the $10spdfg$, $15spdfg$, $20spdfg$, and $25spdfg$ active sets. Here, $nspdfg$ denotes all orbitals up to principal quantum number $n$ for the $s$, $p$, $d$, $f$, and $g$ partial waves.

In the iterative MR expansion, the active set was fixed at $10spdfg$, and the initial reference set $\mathrm{MR}_0$ contained only $6d^3 7s^2$. After each CI diagonalization, the average contribution of a configuration $K$ to the bound states was evaluated as
\begin{equation}
    \bar{w}(K)=
    \frac{1}{N_{\mathrm{b}}}
    \sum_{\alpha=1}^{N_{\mathrm{b}}}
    \sum_{i\in K}
    \left|c_i^{(\alpha)}\right|^2,
    \label{eq:configuration_weight}
\end{equation}
where $N_{\mathrm{b}}$ is the number of bound states and $c_i^{(\alpha)}$ is the CI mixing coefficient of Slater determinant $i$ in bound state $\alpha$. Configurations with average contributions exceeding the selected threshold were added to the MR set, from which SD excitations to $10spdfg$ generated the next CI space. The thresholds were progressively lowered to $0.1$, $0.01$, $0.005$, $0.001$, $0.0005$, and $0.0001$ for $\mathrm{MR}_1$--$\mathrm{MR}_6$, respectively.

\begin{figure}[t]
    \centering
    \includegraphics[width=\columnwidth]{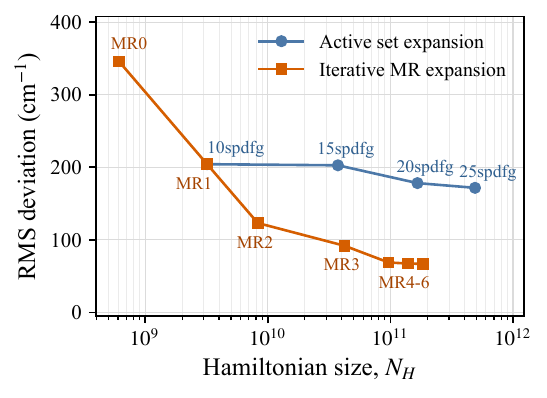}
    \caption{
    Comparison of the two CI space construction strategies for the even-parity states. The RMS deviation is evaluated for the three observed even-parity excitation energies~\cite{Tang2019,Zhang2025}, with the level at $3166.8~\mathrm{cm}^{-1}$ assigned to ${}^4P^e_{3/2}$ as in the main text. The level at $3666~\mathrm{cm}^{-1}$ is excluded because its assignment to ${}^4P^e_{1/2}$ or ${}^4F^e_{9/2}$ remains uncertain.
    }
    \label{fig:sm_ci_convergence}
\end{figure}

Figure~\ref{fig:sm_ci_convergence} compares the root-mean-square (RMS) deviations from the experimental excitation energies~\cite{Tang2019,Zhang2025} and the computational cost, represented by the number of nonzero Hamiltonian matrix elements, $N_H$. For the active set expansion, extending the active set from $10spdfg$ to $25spdfg$ reduces the RMS deviation only from $204.3$ to $171.7~\mathrm{cm}^{-1}$, while increasing $N_H$ by a factor of about 153. In contrast, for the iterative MR expansion, the RMS deviation decreases progressively from $345.7~\mathrm{cm}^{-1}$ for $\mathrm{MR}_0$ to $68.8~\mathrm{cm}^{-1}$ for $\mathrm{MR}_4$, with a difference of only $0.35~\mathrm{cm}^{-1}$ between $\mathrm{MR}_5$ and $\mathrm{MR}_6$. The final $\mathrm{MR}_6$ calculation requires approximately $2.7~\mathrm{TiB}$ for Hamiltonian storage, about $37\%$ of that required by the $25spdfg$ calculation in the active set expansion, while reducing the RMS deviation by $104.8~\mathrm{cm}^{-1}$.
We therefore adopted $\mathrm{MR}_6$ for the final CI+all-order calculations. The MR sets contained 110 even-parity and 113 odd-parity reference configurations. Allowing all possible SD excitations from these sets to $10spdfg$ generated 186,147 even-parity and 208,333 odd-parity relativistic configurations, corresponding to approximately $4.9\times10^{7}$ and $5.0\times10^{7}$ Slater determinants, respectively.

\subsection{MCDHF+RCI calculations}
\label{sec:sm_mcdhf_rci}

For the MCDHF calculations, the initial MR sets consisted of $6d^3 7s^2$, $6d^4 7s$, $6d^2 7s 7p^2$, and $6d^3 7p^2$ for even parity, and $6d^2 7s^2 7p$, $6d^3 7s 7p$, and $6d^4 7p$ for odd parity. Following the orbital optimization procedure described in the main text, we constructed four successive active spaces. The first active space was defined as $\mathrm{AS}_1=\{8s,8p,7d,5f,5g\}$, and the remaining active spaces were defined as $\mathrm{AS}_2=\mathrm{AS}_1+\{9s,9p,8d,6f,6g\}$, $\mathrm{AS}_3=\mathrm{AS}_2+\{10s,10p,9d,7f,7g\}$, and $\mathrm{AS}_4=\mathrm{AS}_3+\{11s,11p,10d,8f,8g\}$. The excitation energies show good convergence with the enlargement of the active space. In particular, extending the active space from $\mathrm{AS}_3$ to $\mathrm{AS}_4$ results in an average absolute change of only $5.6~\mathrm{cm}^{-1}$ in the excitation energies, indicating that the $\mathrm{AS}_3$ results are sufficiently converged. We therefore adopted the corresponding $10sp9d7fg$ active space for the subsequent calculations.

For the RCI calculations, the MR sets were progressively enlarged based on the configuration weights obtained from the preceding MCDHF calculations. The final MR sets comprised 48 even-parity and 43 odd-parity configurations, accounting for approximately $99\%$ of the cumulative configuration weights for each parity. Further enlargement to $99.5\%$ resulted in an average absolute change of only $16.1~\mathrm{cm}^{-1}$ in the excitation energies. To assess the importance of core correlations, correlations involving the $6p$, $6s$, $5d$, and $5p$ subshells were included successively. The corresponding average changes in the excitation energies were $407.4$, $104.4$, $78.6$, and $11.5~\mathrm{cm}^{-1}$, respectively, indicating progressively smaller contributions from the deeper core subshells.

The large RCI expansions were treated using a two-stage zero-first (ZF) procedure~\cite{Li2025Blockwise}. In the first ZF calculation, the zero-order space contained the CSFs describing valence-valence (VV) correlations, while the first-order space accounted for additional core-valence (CV) and core-core (CC) correlations. The resulting mixing coefficients were used to rearrange and condense the CSF(G)s through the CSFG-based condensation procedure~\cite{Li2023Reducing,Si2025Graspg}. The leading CSF(G)s with accumulated weights up to $0.9999$ were used to construct the new zero-order space, whereas those with accumulated weights up to $0.99999$ were retained in the new first-order space. A second ZF calculation with the condensed spaces was then performed to obtain the final MCDHF+RCI results. Increasing the zero- and first-order condensation thresholds to $0.99995$ and $0.999999$, respectively, changed the excitation energies by only a few $\mathrm{cm}^{-1}$, indicating that the adopted thresholds were sufficient to obtain stable excitation energies.

\section{Radiative Transition Properties}
\label{sec:sm_transitions}

Radiative transition properties for electric dipole (E1), magnetic dipole (M1), and electric quadrupole (E2) transitions were calculated using both methods. The electric multipole transition matrix elements were evaluated in the length form. In the CI+all-order calculations, effective operators incorporating random-phase approximation (RPA) corrections were used~\cite{Kozlov2015CIMBPT}. Transition rates $A_{ij}$ were obtained using the theoretical transition energies from the respective methods. The branching fractions and radiative lifetimes were calculated as 
\begin{equation}
    \mathrm{BF}_{ij}
    =
    \frac{A_{ij}}{\sum_k A_{ik}},
    \qquad
    \tau_i
    =
    \left(\sum_k A_{ik}\right)^{-1},
    \label{eq:sm_bf_lifetime}
\end{equation}
where $i$ and $j$ denote the upper and lower levels, respectively, and the sums include all calculated downward radiative channels from level $i$.

Table~\ref{tab:sm_transition_rates} lists the transition rates, branching fractions, and radiative lifetimes obtained by both methods, together with available experimental lifetimes. Transitions with branching fractions below $10^{-7}$ in both calculations are omitted from the table, as their effects are negligible over the typical $10^5$ photon scattering cycles discussed in the main text. 
Overall, the two methods predict very similar decay patterns, with consistent dominant decay channels and comparable branching fractions. Among the 35 transitions listed in Table~\ref{tab:sm_transition_rates}, 31 transition rates agree within a factor of two, indicating good overall consistency. The remaining differences may arise from discrepancies in both the transition matrix elements and the theoretical transition energies. The radiative lifetimes show even closer agreement, with differences within a factor of approximately $1.5$ for all eleven bound excited states.

\begin{table*}[t]
\caption{\label{tab:sm_transition_rates}
Calculated transition rates ($A$), branching fractions (BF), and radiative lifetimes ($\tau$) for the bound states of $\mathrm{Th}^-$, obtained using the CI+all-order and MCDHF+RCI methods, together with available experimental lifetimes. Numbers in brackets denote powers of 10, i.e., $a[b]=a\times10^b$.}
\renewcommand{\arraystretch}{1.15}

\begin{ruledtabular}
\begin{tabular}{ccccccccc}

\multirow{3}{*}{\shortstack{Upper\\level}}
& \multirow{3}{*}{\shortstack{Lower\\level}}
& \multicolumn{2}{c}{BF}
& \multicolumn{2}{c}{$A$ ($\mathrm{s}^{-1}$)}
& \multicolumn{3}{c}{$\tau$} \\
\cline{3-4}\cline{5-6}\cline{7-9}

& & \makecell{CI+\\all-order}
& \makecell{MCDHF\\+RCI}
& \makecell{CI+\\all-order}
& \makecell{MCDHF\\+RCI}
& \makecell{CI+\\all-order}
& \makecell{MCDHF\\+RCI}
& Expt.~\cite{Zhang2025} \\

\hline
\noalign{\vspace{3pt}}

\multicolumn{9}{l}{\textit{Even-parity bound excited states}} \\
\noalign{\vspace{3pt}}

${}^4F^e_{5/2}$
& ${}^4F^e_{3/2}$
& $1.0000$ & $1.0000$
& $1.44[-1]$ & $1.56[-1]$
& $6.96$ s & $6.39$ s & -- \\

\hline
\noalign{\vspace{3pt}}

\multirow{3}{*}{${}^4F^e_{7/2}$}
& ${}^4F^e_{3/2}$
& $2.61[-6]$ & $2.37[-6]$
& $6.03[-5]$ & $4.11[-5]$
& \multirow{3}{*}{$43.3$ ms}
& \multirow{3}{*}{$57.6$ ms}
& \multirow{3}{*}{--} \\

& ${}^4F^e_{5/2}$
& $0.0027$ & $0.0041$
& $6.32[-2]$ & $7.19[-2]$
& & & \\

& ${}^4G^o_{5/2}$
& $0.9973$ & $0.9959$
& $2.31[+1]$ & $1.73[+1]$
& & & \\

\hline
\noalign{\vspace{3pt}}

\multirow{3}{*}{${}^4P^e_{3/2}$}
& ${}^4F^e_{3/2}$
& $0.0862$ & $0.1009$
& $1.84[-1]$ & $1.89[-1]$
& \multirow{3}{*}{$468$ ms}
& \multirow{3}{*}{$535$ ms}
& \multirow{3}{*}{--} \\

& ${}^4F^e_{5/2}$
& $0.0108$ & $0.0111$
& $2.31[-2]$ & $2.08[-2]$
& & & \\

& ${}^4G^o_{5/2}$
& $0.9030$ & $0.8880$
& $1.93[+0]$ & $1.66[+0]$
& & & \\

\hline
\noalign{\vspace{3pt}}

\multirow{2}{*}{${}^4F^e_{9/2}$}
& ${}^4F^e_{5/2}$
& $1.17[-4]$ & $9.55[-5]$
& $1.22[-6]$ & $1.41[-6]$
& \multirow{2}{*}{$95.5$ s}
& \multirow{2}{*}{$67.6$ s}
& \multirow{2}{*}{--} \\

& ${}^4F^e_{7/2}$
& $0.9999$ & $0.9999$
& $1.05[-2]$ & $1.48[-2]$
& & & \\

\hline
\noalign{\vspace{3pt}}

\multirow{4}{*}{${}^4P^e_{1/2}$}
& ${}^4F^e_{3/2}$
& $0.4752$ & $0.5330$
& $3.83[-2]$ & $4.32[-2]$
& \multirow{4}{*}{$12.4$ s}
& \multirow{4}{*}{$12.3$ s}
& \multirow{4}{*}{--} \\

& ${}^4F^e_{5/2}$
& $4.87[-4]$ & $4.14[-4]$
& $3.92[-5]$ & $3.36[-5]$
& & & \\

& ${}^4P^e_{3/2}$
& $0.0204$ & $0.0348$
& $1.64[-3]$ & $2.82[-3]$
& & & \\

& ${}^4F^o_{3/2}$
& $0.5039$ & $0.4317$
& $4.06[-2]$ & $3.50[-2]$
& & & \\

\hline
\noalign{\vspace{3pt}}

\multicolumn{9}{l}{\textit{Odd-parity bound excited states}} \\
\noalign{\vspace{3pt}}

\multirow{2}{*}{${}^4G^o_{5/2}$}
& ${}^4F^e_{3/2}$
& $1.0000$ & $1.0000$
& $7.21[+2]$ & $7.24[+2]$
& \multirow{2}{*}{$1.39$ ms}
& \multirow{2}{*}{$1.38$ ms}
& \multirow{2}{*}{--} \\

& ${}^4F^e_{5/2}$
& $1.14[-6]$ & $7.81[-7]$
& $8.21[-4]$ & $5.66[-4]$
& & & \\

\hline
\noalign{\vspace{3pt}}

\multirow{3}{*}{${}^4F^o_{3/2}$}
& ${}^4F^e_{3/2}$
& $0.9954$ & $0.9934$
& $1.13[+5]$ & $1.04[+5]$
& \multirow{3}{*}{$8.83$ $\mu$s}
& \multirow{3}{*}{$9.55$ $\mu$s}
& \multirow{3}{*}{--} \\

& ${}^4F^e_{5/2}$
& $0.0046$ & $0.0066$
& $5.18[+2]$ & $6.91[+2]$
& & & \\

& ${}^4P^e_{3/2}$
& $3.37[-5]$ & $2.66[-5]$
& $3.82[+0]$ & $2.79[+0]$
& & & \\

\hline
\noalign{\vspace{3pt}}

\multirow{3}{*}{${}^2S^o_{1/2}$}
& ${}^4F^e_{3/2}$
& $0.9935$ & $0.9951$
& $2.40[+4]$ & $2.07[+4]$
& \multirow{3}{*}{$41.3$ $\mu$s}
& \multirow{3}{*}{$48.1$ $\mu$s}
& \multirow{3}{*}{$30(2)$ $\mu$s} \\

& ${}^4P^e_{3/2}$
& $0.0059$ & $0.0047$
& $1.43[+2]$ & $9.68[+1]$
& & & \\

& ${}^4P^e_{1/2}$
& $6.08[-4]$ & $2.24[-4]$
& $1.47[+1]$ & $4.65[+0]$
& & & \\

\hline
\noalign{\vspace{3pt}}

\multirow{4}{*}{${}^4G^o_{7/2}$}
& ${}^4F^e_{5/2}$
& $0.8760$ & $0.9053$
& $4.47[+3]$ & $3.96[+3]$
& \multirow{4}{*}{$196$ $\mu$s}
& \multirow{4}{*}{$229$ $\mu$s}
& \multirow{4}{*}{--} \\

& ${}^4G^o_{5/2}$
& $1.10[-4]$ & $1.25[-4]$
& $5.61[-1]$ & $5.48[-1]$
& & & \\

& ${}^4F^e_{7/2}$
& $0.0888$ & $0.0776$
& $4.53[+2]$ & $3.40[+2]$
& & & \\

& ${}^4F^e_{9/2}$
& $0.0351$ & $0.0170$
& $1.79[+2]$ & $7.43[+1]$
& & & \\

\hline
\noalign{\vspace{3pt}}

\multirow{5}{*}{${}^2D^o_{3/2}$}
& ${}^4F^e_{3/2}$
& $0.2580$ & $0.1568$
& $1.03[+4]$ & $4.10[+3]$
& \multirow{5}{*}{$25.1$ $\mu$s}
& \multirow{5}{*}{$38.2$ $\mu$s}
& \multirow{5}{*}{$62(5)$ $\mu$s\footnotemark[1]} \\

& ${}^4F^e_{5/2}$
& $0.7093$ & $0.8082$
& $2.83[+4]$ & $2.11[+4]$
& & & \\

& ${}^4G^o_{5/2}$
& $1.63[-6]$ & $2.29[-6]$
& $6.50[-2]$ & $6.00[-2]$
& & & \\

& ${}^4P^e_{3/2}$
& $0.0285$ & $0.0321$
& $1.14[+3]$ & $8.39[+2]$
& & & \\

& ${}^4P^e_{1/2}$
& $0.0042$ & $0.0030$
& $1.69[+2]$ & $7.82[+1]$
& & & \\

\hline
\noalign{\vspace{3pt}}

\multirow{5}{*}{${}^2D^o_{5/2}$}
& ${}^4F^e_{3/2}$
& $0.0187$ & $0.0217$
& $2.18[+2]$ & $2.23[+2]$
& \multirow{5}{*}{$85.6$ $\mu$s}
& \multirow{5}{*}{$97.0$ $\mu$s}
& \multirow{5}{*}{$53(3)$ $\mu$s\footnotemark[1]} \\

& ${}^4F^e_{5/2}$
& $0.5802$ & $0.6939$
& $6.78[+3]$ & $7.15[+3]$
& & & \\

& ${}^4G^o_{5/2}$
& $2.09[-5]$ & $2.15[-5]$
& $2.45[-1]$ & $2.21[-1]$
& & & \\

& ${}^4F^e_{7/2}$
& $0.3346$ & $0.2435$
& $3.91[+3]$ & $2.51[+3]$
& & & \\

& ${}^4P^e_{3/2}$
& $0.0665$ & $0.0409$
& $7.77[+2]$ & $4.22[+2]$
& & & \\

\end{tabular}
\end{ruledtabular}

\footnotetext[1]{These experimental lifetimes were originally reported for the resonances tentatively assigned to the $^4F^o_{5/2}$ and $^4D^o_{1/2}$ states in Ref.~\cite{Zhang2025}, respectively. Here, they are associated with the $^2D^o_{3/2}$ and $^2D^o_{5/2}$ states, respectively, according to the reassignments discussed in the main text.}

\end{table*}

\section{Analysis of Laser Cooling Schemes}
\label{sec:sm_cooling}

Here we provide the formulas used to estimate the Doppler cooling limits, required number of photon scattering cycles, and ideal cooling times reported in the main text. For each cooling scheme, the Doppler cooling limit is given by
\begin{equation}
    T_D=\frac{\hbar\Gamma}{2k_B},
    \label{eq:sm_doppler}
\end{equation}
where $\Gamma=\sum_j A_{uj}$ is the total radiative decay rate of the upper level, and $k_B$ is the Boltzmann constant. The calculated Doppler cooling limits are $0.092/0.079~\mu\mathrm{K}$ for Scheme I, based on the ${}^4F^e_{3/2}\leftrightarrow{}^2S^o_{1/2}$ transition, and $0.43/0.40~\mu\mathrm{K}$ for Scheme II, based on the ${}^4F^e_{3/2}\leftrightarrow{}^4F^o_{3/2}$ transition. Paired values correspond to the CI+all-order and MCDHF+RCI results, respectively.

For one-dimensional cooling from an initial temperature $T_i$ to a final temperature $T_f$, the number of photon scattering cycles required for the corresponding momentum change is estimated as
\begin{equation}
    N_\gamma \simeq
    \frac{M}{\hbar k}
    \left[
    \sqrt{\frac{k_B T_i}{M}}
    -
    \sqrt{\frac{k_B T_f}{M}}
    \right],
    \label{eq:sm_photon_number}
\end{equation}
where $M$ is the mass of the \ce{Th^-} ion and $k=2\pi/\lambda$ is the wave number of the cooling transition. Taking $T_i=100~\mathrm{K}$ and $T_f=T_D$, we obtain approximately $8.7\times10^4$ scattering cycles for Scheme I and $9.7\times10^4/9.5\times10^4$ cycles for Scheme II.

Under ideal two-level saturation conditions, the photon scattering rate approaches $R_{\mathrm{sat}}\simeq A_{ug}/2$, where $A_{ug}$ is the spontaneous transition rate from the upper cooling state to the ground state. The corresponding ideal cooling time is
\begin{equation}
    t_{\mathrm{ideal}}
    =\frac{N_\gamma}{R_{\mathrm{sat}}}
    \simeq\frac{2N_\gamma}{A_{ug}},
    \label{eq:sm_ideal_time}
\end{equation}
giving $7.2/8.4~\mathrm{s}$ for Scheme I and $1.7/1.8~\mathrm{s}$ for Scheme II.

Both cooling schemes exhibit population leakage to metastable states. To assess its effect, we first estimate the cumulative shelving time in each state $m$ as
\begin{equation}
    t_m \simeq N_\gamma\mathrm{BF}_{um}\tau_m,
    \label{eq:sm_shelving_estimate}
\end{equation}
where $\mathrm{BF}_{um}$ is the branching fraction from the upper cooling state $u$ to state $m$, and $\tau_m$ is its radiative lifetime. Without repumping, $N_\gamma$ photon scattering cycles result in approximately $N_\gamma\mathrm{BF}_{um}$ leakage events, each leading to an average shelving time of $\tau_m$. The resulting $t_m$ allows a qualitative assessment of the effect of each leakage channel on cooling. As shown in Table~\ref{tab:sm_shelving_times}, the estimated shelving times for ${}^4P^e_{1/2}$ and ${}^4P^e_{3/2}$ in Scheme I and ${}^4F^e_{5/2}$ in Scheme II range from hundreds to thousands of seconds, far exceeding the ideal cooling times. Repumping these states is therefore necessary for cooling on the timescales considered here. In contrast, the estimated shelving time for ${}^4P^e_{3/2}$ in Scheme II is only $1.54/1.35~\mathrm{s}$, owing to its much smaller leakage branching fraction. This shelving time is substantially shorter than those of the other leakage channels but remains comparable to the ideal cooling time of Scheme II. A more detailed analysis is therefore needed to assess the impact of repumping this state.

\begin{table}[t]
\caption{\label{tab:sm_shelving_times}
Estimated cumulative shelving times $t_m$ for metastable states $m$.
Here, $\mathrm{BF}_{um}$ is the branching fraction from the upper
cooling state $u$ to state $m$, and $\tau_m$ is the radiative lifetime
of state $m$. Paired values denote the CI+all-order and MCDHF+RCI
results, respectively. Numbers in brackets denote powers of ten.
}
\setlength{\tabcolsep}{3pt}
\begin{ruledtabular}
\begin{tabular}{cccc}
State & $\mathrm{BF}_{um}$ & $\tau_m$ (s) & $t_m$ (s) \\
\hline
\noalign{\vspace{3pt}}
\multicolumn{4}{l}{\textit{Scheme I}} \\
\noalign{\vspace{3pt}}

${}^4P^e_{1/2}$
& $6.08[-4]/2.24[-4]$ & $12.4/12.3$ & $654/239$ \\
${}^4P^e_{3/2}$
& $5.89[-3]/4.66[-3]$ & $0.468/0.535$ & $239/216$ \\

\hline
\noalign{\vspace{3pt}}
\multicolumn{4}{l}{\textit{Scheme II}} \\
\noalign{\vspace{3pt}}

${}^4F^e_{5/2}$
& $4.58[-3]/6.60[-3]$ & $6.96/6.39$ & $3108/4007$ \\
${}^4P^e_{3/2}$
& $3.37[-5]/2.66[-5]$ & $0.468/0.535$ & $1.54/1.35$ \\

\end{tabular}
\end{ruledtabular}
\end{table}

To quantify this effect, we constructed a multilevel rate equation model including all twelve bound states, spontaneous decay, laser excitation, and stimulated emission. The population $P_i$ of state $i$ evolves according to
\begin{equation}
    \frac{dP_i}{dt}
    =
    \sum_{j\ne i}
    (K_{ji}P_j-K_{ij}P_i),
    \label{eq:sm_rate_equation}
\end{equation}
where $K_{ij}$ is the total transition rate from state $i$ to state $j$, including spontaneous decay and laser driven contributions. The spontaneous decay rates were taken from Table~\ref{tab:sm_transition_rates}. For a laser driven transition between the lower level $a$ and upper level $b$, the laser excitation and stimulated emission rates $W_{ab}$ and $W_{ba}$ satisfy
\begin{equation}
    W_{ba}
    =
    \frac{2J_a+1}{2J_b+1}W_{ab}.
    \label{eq:sm_laser_rates}
\end{equation}

For Scheme II, the laser excitation rates of the cooling and ${}^4F^e_{5/2}$ repumping transitions were both set to $10^7~\mathrm{s}^{-1}$ to represent strong driving. Two cases were considered, with and without additional repumping of ${}^4P^e_{3/2}$, using the same laser excitation rate when repumping was applied. The steady state was obtained by setting $dP_i/dt=0$ and $\sum_iP_i=1$. Hereafter, $P_i$ denotes the steady state population. The effective photon scattering rate is
\begin{equation}
    R_{\mathrm{eff}}
    =
    P_gW_{gu}-P_uW_{ug},
    \label{eq:sm_scattering_rate}
\end{equation}
where $g$ and $u$ denote the ground and upper cooling states, respectively. The two terms represent laser excitation and stimulated emission rates, respectively. Assuming that the steady state is reached rapidly and $R_{\mathrm{eff}}$ remains approximately constant during cooling, the cooling time is estimated as
\begin{equation}
    t_{\mathrm{cool}}
    \simeq\frac{N_\gamma}{R_{\mathrm{eff}}}.
    \label{eq:sm_cooling_time}
\end{equation}

\begin{table}[t]
\caption{\label{tab:sm_rate_results}
Steady state results from the multilevel rate equation calculations for Scheme II with (On) and without (Off) additional repumping of ${}^4P^e_{3/2}$. Both cases include repumping of ${}^4F^e_{5/2}$. Here, $P({}^4P^e_{3/2})$ is the steady state population of ${}^4P^e_{3/2}$, $R_{\mathrm{eff}}$ is the effective photon scattering rate, and $t_{\mathrm{cool}}$ is the estimated cooling time. Paired values denote the CI+all-order and MCDHF+RCI results, respectively.
}
\renewcommand{\arraystretch}{1.25}
\begin{ruledtabular}
\begin{tabular}{lccc}
Repumping
& $P({}^4P^e_{3/2})$
& $R_{\mathrm{eff}}$ ($10^4~\mathrm{s}^{-1}$)
& $t_{\mathrm{cool}}$ (s) \\
\hline
Off
& $0.337/0.298$
& $2.12/2.08$
& $4.59/4.57$ \\
On
& $0.222/0.222$
& $2.50/2.31$
& $3.91/4.12$
\end{tabular}
\end{ruledtabular}
\end{table}

The results of the multilevel rate equation calculations presented in Table~\ref{tab:sm_rate_results} show that additional repumping of ${}^4P^e_{3/2}$ reduces the estimated cooling time by only $14.9\%$ and $9.8\%$ for the CI+all-order and MCDHF+RCI results, respectively. These modest improvements may not justify the additional experimental complexity of another repumping laser, suggesting that repumping this state is not essential for Scheme II.

\bibliography{ref}